\documentclass[12pt, letterpaper]{article}
\usepackage[top=1in, bottom=1.5in, left=1in, right=1in]{geometry}
\usepackage{authblk}
\usepackage[utf8]{inputenc}
\usepackage{booktabs}
\usepackage{hyperref}
\usepackage{graphicx}
\usepackage{natbib}
\usepackage{tabularx,calc}
\usepackage{epsfig,color}
\usepackage{rotating}
\usepackage{multicol}
\usepackage{amssymb}
\usepackage{amsmath}
\usepackage{microtype}
\usepackage{enumitem}
\usepackage{floatrow}

\title{LSST Cadence Optimization White Paper in Support of Observations of Unresolved Tidal Stellar Streams in Galaxies beyond the Local Group}
\author{Seppo Laine}
\affil{Caltech/IPAC}
\author{David Mart\'{i}nez-Delgado}
\affil{ARI/U. Heidelberg}
\author{Ignacio Trujillo}
\affil{Instituto de Astrof\'{i}sica de Canarias}
\author{Pierre--Alain Duc}
\affil{ObAS, U. Strasbourg}
\author{Carl J. Grillmair}
\affil{Caltech/IPAC}
\author{Carlos S. Frenk}
\affil{U. Durham, UK}
\author{David Hendel}
\affil{U. Toronto}
\author{Kathryn V. Johnston}
\affil{Columbia U.}
\author{J. Chris Mihos}
\affil{CWRU}
\author{John Moustakas}
\affil{Siena College}
\author{Rachael L. Beaton}
\affil{Princeton University}
\author{Aaron J. Romanowsky}
\affil{UC Obs., SJSU}
\author{Johnny Greco}
\affil{CCAP, The Ohio State University}
\author{Denis Erkal}
\affil{University of Surrey}

\date{November 30, 2018}

\begin{document}

\maketitle

\begin{abstract}

Deep observations of faint surface brightness stellar tidal streams in external  galaxies
with LSST are addressed in this White Paper contribution. We propose using the
Wide--Fast--Deep survey that contains several nearby galaxies (at distances where the stars
themselves are not resolved, i.e., beyond 20 Mpc). In the context of hierarchical galaxy
formation, it is necessary to understand the  prevalence and properties of tidal
substructure around external galaxies based on integrated (i.e., unresolved) diffuse light.
This requires collecting observations on much larger samples of galaxies than the Milky Way
and M31. We will compare the observed structures to the predictions of cosmological models
of galactic halo formation that inform us about the number and properties of streams around Milky Way-like galaxies. The insight gained from these comparisons will allow us to infer the properties of stream progenitors (masses, dynamics, metallicities, stellar
populations). The changes in the host galaxies caused by the interactions with the
dissolving companion galaxies will be another focus of our studies. We conclude by discussing
synergies with WFIRST and Euclid, and also provide concrete suggestions for how the effects
of scattered light could be minimized in LSST images to optimize the search for low surface
brightness features, such as faint unresolved stellar tidal streams.

\end{abstract}

\section{White Paper Information}
Please contact Seppo Laine, Caltech/IPAC, seppo@ipac.caltech.edu, with 
any questions about this white paper. 

This white paper addresses:
\begin{enumerate} 
\item {\bf Science Category:} Milky Way Structure and Formation: Exploring the Faint Surface Brightness Universe.
\item {\bf Survey Type Category:} Wide--Fast--Deep Survey.
\item {\bf Observing Strategy Category:} Targeting areas of sky off the Galactic plane.
\end{enumerate}  

\clearpage

\section{Scientific Motivation}

Previous deep, wide-area photometric surveys have revealed a large number
of faint stellar substructures (``streams'') around galaxies like the Milky Way (MW),
resulting from the tidal disruption of lower-mass galaxies in ``minor mergers'' or previous
major mergers. While detailed studies of resolved streams around the MW and M31 imply a
dynamic hierarchical accretion history, consistent with $\Lambda$CDM cosmological galaxy
formation models
\citep[e.g.,][]{bullock05,delucia08,cooper10,cooper13,pillepich15,rodriguez16}, we need
to study a much larger sample of galaxies to test whether the merging histories of MW and
M31 are typical of galaxies in their mass range \citep[e.g.,][]{mutch11,morales18}.

A crucial ingredient in testing whether the merging histories of the MW and M31 are
typical (consistent with $\Lambda$CDM cosmological galaxy formation models) is the
acquisition of  adequately deep images, as the majority of the predicted tidal stellar
streams have surface brightnesses in the R-band fainter than about 29 AB mag arcsec$^{-2}$.
While a few deep imaging surveys of the outskirts of local galaxies have recently been
completed \citep[e.g.,][]{tal09,delgado10,ludwig12,duc15} or are ongoing \citep{abraham14},
the majority of nearby galaxies have not been observed down to surface brightnesses needed
to detect streams from ancient minor mergers. 

By focusing on nearby spiral galaxies with diffuse-light overdensities, more than 50
previously unknown stellar structures in galaxies at distances $<$ 80 Mpc have been
discovered so far (Mart\'{i}nez-Delgado 2018). The morphologies of the diffuse-light structures include ``great circle''-like
(such as are seen around the Milky Way) streams that roughly trace the orbit of the merging satellite galaxy \citep{sanders13}, isolated shells, giant debris clouds,
jet-like features, and large diffuse structures that may be old, phase mixed remnants of
merged companions (see Figure \ref{fig:decals} for examples). Again, very similar features
are seen in cosmological simulations of minor mergers \citep{johnston08,cooper10}. Although
it appears in many cases that the progenitor companion has been completely disrupted, a few
examples \citep{delgado12,delgado15,amorisco15} show surviving cores of merging companion
galaxies, often exhibiting long tails departing from the progenitor satellite.

\begin{figure}[t]
\centering
   \epsfig{figure=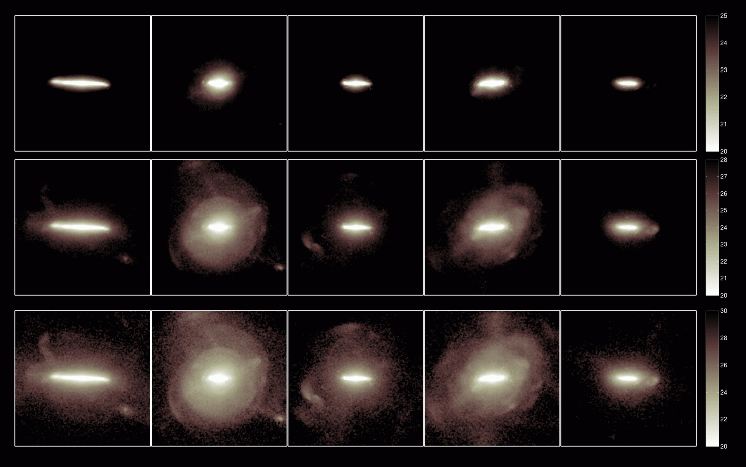, height=8.2cm, width=12cm}
    \caption{\looseness=-1 \footnotesize{Expected `halo streams' around an MW-like
galaxy from the Auriga cosmological simulations \citep{grand17}. The panels
show an external perspective of several realizations of a simulated galaxy within the hierarchical framework, with streams resulting from tidally 
disrupted satellites. They illustrate a variety of typical accretion histories for MW-like galaxies. Each panel is 300~kpc on a side. The different rows show theoretical  predictions for detectable tidal features in each halo model, assuming three different surface brightness (SB) detection limits (bottom row: $\mu_{\mathrm{lim}}=31$, middle row: $\mu_{\mathrm{lim}} = 28$ and top row: $\mu_{\mathrm{lim}}= 25$ AB mag/arcsec$^2$). %Each {\it snapshot} is 100 kpc on a side, and correspond to the typical halo area covered by our survey in each target.
This suggests that the number of tidal features visible in the outskirts of spirals varies dramatically with the SB limit of the data, with no discernible sub-structure expected for surveys with SB limits brighter than $\sim25$ AB mag/arcsec$^2$ (e.g.\ POSS-II and SDSS). For the expected SB limit of the LSST ($\sim31$ AB mag/arcsec$^2$ in $g$-band; see Sec. 4), we would expect to detect streams around $\sim$ 80--90\% of our galaxy sample.}}
    \label{zoo}
\end{figure}

One of the main objectives of a conceivable LSST stream survey among MW-like galaxies
is the comparison of the observed frequency and the parameters of the streams, such as their
spatial coherence, length, width, inclination, morphology, color and surface brightness to
those seen in cosmological simulations. These models suggest that remnants from mergers 0--8
Gyrs ago are still visible as substructures in the halos of nearby galaxies. Information
about the streams comes from 1) morphology of the low surface brightness emission, as
\citet[][their Fig. 3]{johnston08} and \citet{hendel15} have shown that different morphologies of
tidal debris occupy different regions in the accretion time vs. orbital eccentricity/energy
plane; 2) the progenitor's luminosity; 3) stellar population of the stream (to infer its
stellar mass); 4) shape and width of the streams \citep[to obtain the dynamical properties of the
progenitors;][]{johnston01,erkal16}; 5) color of the stream (to determine whether a minor or major merger was
involved); 6) mass and morphological type of the host galaxy (to obtain the frequency of
mergers); and 7) (combined with the  previously listed information) surface brightness that
can be used to time the epoch of the merger \citep[and thus the rate at which new streams
are being formed in the local Universe;][]{johnston01}.

The halos in the simulations of \citet{bullock05} typically have about two streams brighter
than 30 AB mag~arcsec$^{-2}$. However, the majority of substructures are at surface
brightnesses fainter than 30 AB mag~arcsec$^{-2}$. Our current inability to see the fainter
streams (corresponding to either earlier merger epochs or lower mass progenitors) implies
that currently the merger history that we study in galaxies beyond the Local Group is
strongly biased towards the most recent (the last few tens of percent of mass accretion)
and/or most massive minor merger events. Thus we are only sensitive to the most metal-rich populations, as
validated by studies of resolved stars around M31 \citep{mcconnachie18} and Cen A
\citep{crnojevic16}. Close to the center of a galaxy (R$_{\rm proj}$ $<$ 30 kpc) the
substructure is most likely generated by the most massive merged satellite galaxies, as
dynamical friction will have brought them quickly to the central regions. Therefore, our
current view of tidal streams in nearby galaxies is highly biased towards the most massive
minor mergers which are relatively rare for MW-like galaxies.

Studying the frequency, mass ratios and stellar populations of minor mergers with the help
of stellar tidal streams in external galaxies will also provide critical input on a number
of open astrophysical questions, such as 1) heating and thickening of the host galaxy
disk by satellite mergers, including the frequency and impact of  low orbital inclination
satellites, most recently showcased by the Gaia-Enceladus galaxy remains in the MW
\citep{helmi17}; 2) as discussed above, the hierarchical build-up of the primaries;
3) dark matter halo shapes, when combined with N-body simulations and 4) the fundamental
nature of  dark matter. Regarding this last point, there are alternative candidate
particles to cold dark matter that are well motivated from particle physics and that result
in radically different properties for astronomical objects on the small scales probed by
stellar streams. One of the currently popular alternatives to CDM is sterile neutrinos with
a mass of a few keV, which behave as warm dark matter (WDM). On the large scales probed by
the cosmic microwave background and large-scale structure, they are indistinguishable from
CDM. But on small scales they are very different. In particular, they require a cutoff in
the mass function of dark matter halos on scales of about 10$^{9}$ M$_{\odot}$ for particle
masses of interest. It is likely that WDM models predict different properties for tidal
stellar streams, although this is still to be verified with high-resolution simulations.
Other particle models, such as certain types of self-interacting particles, also predict a
cutoff in the halo mass function, which may also lead to differences in the properties of
the streams.

The spectral energy distributions (SEDs) of the merged galaxies can be best studied by
combining the LSST observations with data from near-IR wavelengths. WFIRST will
provide a survey of several deep fields efficiently and will provide an ideal
complementary data set near the peak of the stellar SED. The optical images from Euclid may also be useful. The combined LSST/Euclid/WFIRST data set will provide a broad
wavelength baseline for the estimation of the ages, metallicities and masses of the
stellar populations of disrupted companions, producing valuable constraints on the minor
merging history in CDM models of hierarchical galaxy formation.

\begin{figure}[t]
\includegraphics[width=1.0\textwidth]{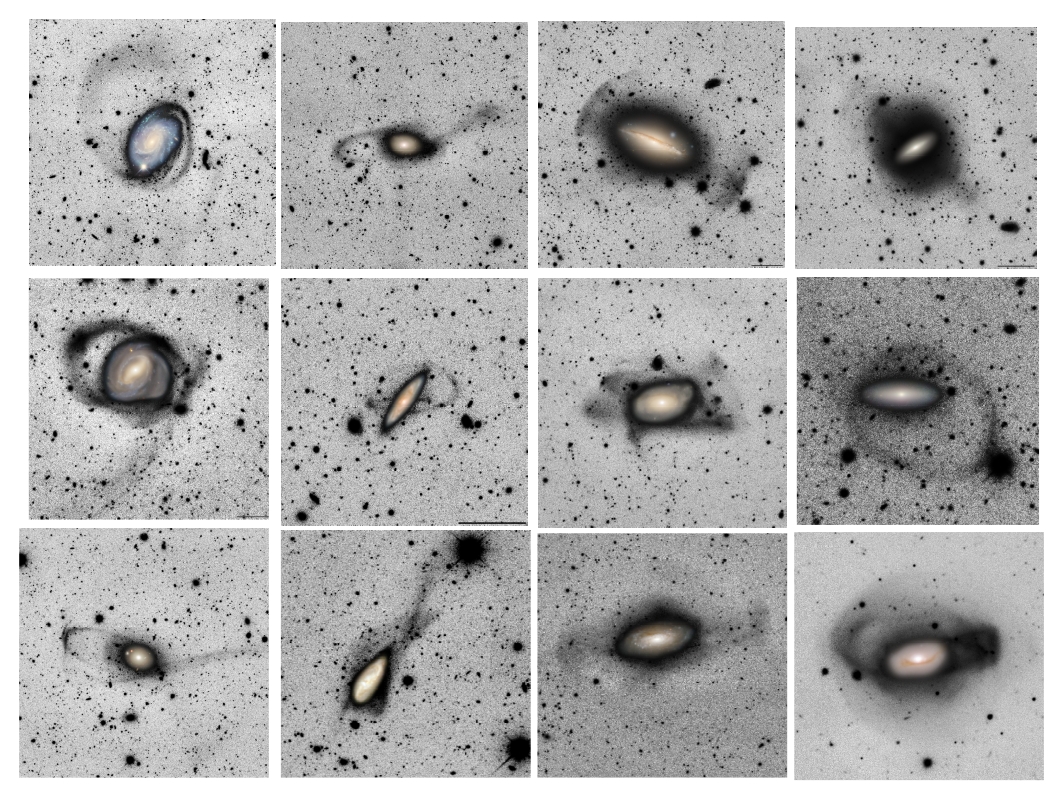}
\caption{DECaLs stacked image cutouts. The distance range of these galaxies is 
30--100 Mpc. Color insets of the central region of the host galaxies have been added to the negative version of the images (Mart\'{i}nez--Delgado et al., in preparation).
\label{fig:decals}}
\end{figure}

\clearpage

\section{Technical Description}

\subsection{High-level description}

Our project will be implemented in essentially three steps: 1) producing deep images from
the Wide--Fast--Deep (WFD) survey with LSST and combining the data from WFIRST/Euclid; 2)
searching systematically (both by visual inspection and by automatic detection algorithms
trained by visual inspection detections) for tidal stellar streams in integrated light and
quantifying their parameters and frequency, together with detection limits; and 3)
quantitatively comparing the observations to results from cosmological simulations (mock
images from the latter will be created, including realistic observational artifacts and
systematic errors).

Detecting the faint streams requires dark-sky conditions and high precision calibration data
(e.g., exquisite flat-field quality over a relatively large angular scale). More
specifically, stellar streams are typically found at large galactocentric distances (15 kpc
$<$ R $<$ 100 kpc, or farther) and could be found out to a significant portion of the virial
radius of the parent galaxy (for the MW or M31, R$_{\rm virial}$ $\lesssim$ 250 kpc).  Thus,
surveys for stellar debris must produce images over large angular scales (from  $>$ 50$'$
for systems at $D$ $\sim$~20 Mpc). Our requirement for resolving the widths of the streams
is 200 pc (2$''$ -- 0.8$''$ at 20 -- 50 Mpc; the median seeing requirement for LSST is
0.7$''$), as we are interested in studying streams left behind by dwarf galaxies (the
globular cluster streams in our Galaxy are about 100 pc in width). However, in practice we
need much higher resolution than 2 arcseconds as we need to resolve and mask out the
background galaxies (complementary higher resolution WFIRST and Euclid observations, discussed later, will help). We also aim to locate the progenitor along the stream, if it has not
been completely disrupted. The various bands of the LSST are needed to study color
variations along the stream. The mean color of the stream will be calculated by averaging
luminosities from the LSST bands that have sufficient depth. All color information will be
used to constrain the spectral energy distribution and  compared to those seen in
cosmological simulations of streams. The deepest band can be used to set the apertures where
the colors are measured. Finally, proper SED modeling can be done to derive stellar masses
\citep[marginalizing over the uncertainties in metallicity, dust and star formation history;
e.g.,][]{zibetti09,laine16}.

\subsection{Sample}

Because we are interested in nearby galaxies (streams at larger distances than usual can be
imaged due to the smaller PSF core size of the LSST than in many past and current surveys), 
and isolated systems in order to exclude major interactions and  mergers from our sample, we
will use the data from the main WFD. We will select mostly galaxies that
are ``analogs to the MW.'' This will help us to compare the upcoming detailed
observations of the MW with LSST and Gaia with similar galaxies in the nearby
Universe, allowing us to estimate whether the Milky Way is typical in its accretion history.
Therefore, we will select galaxies with an absolute $K$-band AB magnitude of $-19.6$ or
brighter. We will also select  galaxies that are away from the Galactic plane,
$\vert$b$\vert$ $>$ 20$^{\circ}$, to avoid confusion with cirrus. While Galactic cirrus emission can be found in every direction around the MW, we will compare the LSST
images of diffuse faint surface brightness emission to images from WISE, AKARI and IRAS (and
Herschel, when available) around 100 microns to avoid misidentifying diffuse emission with
cirrus \citep{mihos17}. We will also impose isolation criteria on our sample galaxies, such
that isolated galaxies  and galaxies in ``fossil groups'' are included (those where the
difference between the brightest and second brightest members is larger than 2.5 mag inside a
projected radius of 1 Mpc and  $\vert$V$_{gal}$ $-$ V$_{neighbor}$$\vert$ $<$ 250
km~s$^{-1}$; \citeauthor{karachentsev09} \citeyear{karachentsev09}).

Mock images made out of cosmological simulations will include realistic observational
effects such as sky noise, flat-field uncertainties, and contamination from background and
foreground objects for the comparison with LSST image stacks. The models include
magneto-hydrodynamical simulations from the IllustrisTNG project 
\citep{weinberger17,pillepich18} and the Copernicus Complexio ({\it CoCo}) cosmological 
N-body simulations \citep{hellwing16}.

\subsection{Image depth}

We target an image depth $>$ 29 AB mag~arcsec$^{-2}$. Expected sky brightness is $\sim$~22 V AB
mag~arcsec$^{-2}$.

\subsection{Filter choice}

A broad wavelength coverage from 0.3 to 1.1 $\mu$m (LSST filters $u$, $g$, $r$, $i$, $z$,
$y$) is optimal for diagnostics  of stellar populations. Our ability to obtain colors will
be limited by the filter with the shallowest data we can use, and therefore we will focus
on $ugr$ imaging. We plan to extend the wavelength coverage to 2 microns using WFIRST images
in the areas where those data are available. 

\subsection{Ideal pointing}

The whole WFD survey area outside the Galactic plane ($\vert$b$\vert$ $>$ 20$^{\circ}$).

\subsection{Exposure constraints}

No constraint on exposure length. The proposed 2$\times$15 sec visits should work well
for a tidal stellar stream survey, as a large number of visits can be used to reject variable
phenomena (both image artifacts and astrophysical events) from the images.

\subsection{Other constraints}

While low surface brightness observations can benefit from dithering and sky rotation to eliminate scattered light from the individual exposures, what really matters is the background subtraction. A local background will eliminate all 
extended LSB features. For NGVS (The Next Generation Virgo Cluster Survey) and MATLAS (Mass 
Assembly of early-Type GaLAxies with their fine Structures) the background was determined
from either  adjacent fields or images obtained with large offsets, larger than the size of
the  structure we want to probe. This allowed us to produce a large background image that could be 
subtracted from all individual ones. The gain was 2--3 mags with such a strategy, as it 
eliminates systematic effects in the camera. We want to make sure that the LSST will allow the subtraction of a large background area and that the pipeline will not eliminate what we are seeking.

Reflections and scattered light in the optical path are a real concern for deep wide-field
surface photometry. While the LSST has spent significant effort to minimize scattered
light, reflections from bright stars in the field can still contaminate images with extended
diffuse halos of light. Worse yet, these reflections are not static; they move with respect
to their parent star, depending on the position of the star in the optical beam. {\it This
means that static star subtraction models applied to stacked images will be insufficient to
remove these features}. Instead, active-subtraction techniques applied at the data reduction
stage \citep[e.g.,][]{slater09} must be used to deal with these reflections by modeling and
removing them on a star-by-star basis from the individual raw image frames. {\bf This makes
it imperative that LSST data servers provide users with the raw images, not just the image
stacks.} 

\subsection{Technical trades}

We would prefer deeper observations at the expense of covering a smaller fraction of
the sky, as our sample size is expected to be sufficient for statistically significant
conclusions even in the case of reduced sample numbers compared to what was stated
above.

The trade-off between the single visit exposure time and the number of visits is not relevant to the proposed science, unless the number of visits drops to ten or fewer, which is extremely unlikely.

The single visit limiting depth is not of concern to us either. The overall uniformity
of depth over the survey field is more important. If this depth varies by factors of
two or more, it is more difficult to obtain statistically significant conclusions for
a large sample size.

It would be acceptable to drop the depth in two or three bands to substantially below 
the rest of the bands. We need at least three deep photometric bands (preferably spaced
as far from each other in wavelength as possible) to perform meaningful estimates of the stellar
populations in any detected streams.

\subsection{Combining LSST data with WFIRST and Euclid}

To obtain the most accurate possible determination of the stellar populations of the
detected streams, we will combine the LSST images with images from WFIRST and Euclid.
These will extend the coverage further into the near-infrared and produce more accurate
determinations of the stellar populations, but the common survey areas are smaller than 
the LSST WFD survey (about 11,000 square degrees with Euclid and much smaller with
WFIRST). Different possible scenarios of combining the data from these missions are
under consideration, including pixel-matched measurement and separate measurements with
native pixel size but over the exact same sky region.

\section{Performance Evaluation}

The most obvious heuristic is the total depth achieved during the survey which is directly related to the number of detections of low surface brightness streams in integrated light. Based on the deepest imaging observations ever taken, 8 hours on-source using the 10.4-meter Grant Telescopio de Canarias (GTC) telescope and with an average seeing of 0.8--0.9 arcseconds \citep{trujillo16}, we have made an approximate estimate of the expected low surface brightness limit that the LSST can provide after the scheduled 825 visits to the same sky location. This corresponds to a total amount of time on source of 3.44h. The expected surface brightness limits will be (3$\sigma$; 10$\times$10 arcsec$^{2}$ boxes): 29.9 ($u$), 31.1 ($g$), 30.6 ($r$), 30.1 ($i$), 28.7 ($z$) AB mag arcsec$^{-2}$. Each visit that consists of 30 seconds on-source will correspond to the following limits (3$\sigma$): 26.6 ($u$), 27.8 ($g$), 27.3 ($r$), 26.8 ($i$), 25.4 ($z$) AB mag arcsec$^{-2}$. If the
WFD survey depth is increased by 10\%, our total detection limit goes up only by 0.05 mag arcsec$^{-2}$.

%The current WFD survey would produce depths of about XXX AB~mag  arcsec$^{-2}$~pixel$^{-1}$
%in bands $ugrizy$. With this depth and by combining all  the six photometric bands, we
%expect to obtain a total 5$\sigma$ detection depth  of XXX AB~mag
%arcsec$^{-2}$~pixel$^{-1}$,

As suggested by the state-of-the art cosmological simulations (e.g., see Figure 1), these surface brightness limits would enable us to detect new stellar tidal streams in diffuse or integrated light around 80--90\% of our sample of a few thousand galaxies out to several hundred Mpc. % If the
%WFD survey depth is increased by 10\%, our total detection number goes up by XXX.
For obtaining stellar population diagnostics, we require at least a 5$\sigma$
detection in a minimum of three bands. We estimate that with the current WFD survey
parameters we would be able to perform stellar population diagnostics in a few thousand stellar tidal stream systems in integrated light. % If the depth can be increased by 10\%, our
%estimate for the number of systems with adequate colors goes up by XXX.

\section{Special Data Processing}

Obtaining quantitative information from measurements of low surface brightness emission
requires reducing data in a completely different way from that of point source driven
data. One requirement we impose on the data is that individual images should be made
available, not just coadds, as scattered light is much easier to remove from individual 
frames than from image stacks. In addition to the removal of scattered light, there are
several other considerations for low surface brightness measurements. While the LSST data
reduction pipeline masks high surface brightness sources, the effects from these sources 
and bright stars can masquerade as low surface brightness emission because of the
extended PSF. We propose to use software such as {\sc imfit} \citep{erwin15} to remove
the stellar envelope from bright host galaxy emission by modeling the host galaxy
surface brightness model assuming it is a simple Sersic profile.

We will also select images for the deepest image stacks based on sky brightness at the
time of the observations, taking into account the vicinity (and phase) of the Moon to the observed area of the sky, as well as experiment with using all data, regardless of the Moon phase.

Another consideration for the detection of low surface brightness emission is stellar
aureole (from ice particles in thin cirrus clouds in the  atmosphere). We propose to
develop techniques to remove this emission from the LSST images.

To estimate the limiting surface brightness in the image stacks, we intend to run
simulations by injecting substructure into the simulated images to test recovery as a function of surface brightness.

Once the images have been reduced, and the erroneous sources of low surface brightness
emission have been removed, we will use special techniques to  detect tidal stellar streams
in diffuse (or integrated) light, such as software adapted from HSC--SSP \citep[Subaru Hyper
Suprime-Cam Strategic Program;][]{kado-fong18}. We will use adaptive smoothing techniques
\citep[similar to][]{zibetti10}, NoiseChisel and iterative unsharp masking techniques
that separate structure in the image by spatial frequency. While we will inspect the images
by eye to detect features, we will also develop algorithms for automatic detection and
morphological classification of streams, which are expected to be available by the time LSST starts operations \citep[c.f.][]{hendel18}.

Finally, we will perform an estimation of the limiting surface brightness (3$\sigma$)
in the stacked images and will make the re-reduced stacked deep images available
to the community as soon as the basic image reduction has been completed.

\end{document}